\documentclass[12pt]{iopart}
\usepackage{graphicx}
\usepackage{mydefinitions}
\usepackage{color}
\usepackage{enumerate}

\begin{document}
\title{Remote frequency measurement of the ${^1\!S_0}\rightarrow{^3\!P_1}$ transition in laser cooled $^{24}$Mg}
\author{J Friebe$^1$, M Riedmann$^1$, T W\"ubbena$^1$, A Pape$^1$, H Kelkar$^1$, W Ertmer$^1$, O Terra$^2$, U Sterr$^2$, S Weyers$^2$, G Grosche$^2$, H Schnatz$^2$, and E M Rasel$^1$}
\address{$^1$ Institut f\"ur Quantenoptik, Leibniz {Universit\"at} Hannover, Welfengarten 1, 30167 Hannover, Germany}
\address{$^2$ Physikalisch-Technische Bundesanstalt, Bundesallee 100, 38116 Braunschweig, Germany}
\ead{rasel@iqo.uni-hannover.de}

\begin{abstract}
We perform Ramsey-Bord\'e spectroscopy on laser cooled magnesium atoms in free fall to measure the ${^1\!S_0}\rightarrow{^3\!P_1}$ intercombination transition frequency. The measured value of 655 659 923 839 730 (48) Hz is consistent with our former atomic beam measurement \cite{Friebe08}. We improve upon the fractional accuracy of the previous measurement by more than an order of magnitude to $7\times10^{-14}$.
The magnesium frequency standard was referenced to a fountain clock of the Physikalisch Technische Bundesanstalt (PTB) via a phase-stabilized telecom fiber link and its stability characterized for interrogation times up to 8000~s. The high temperature of the atomic ensemble leads to a systematic shift due to the motion of atoms across the spectroscopy beams. In our regime, this leads to a counterintuitive \emph{reduction} of residual Doppler shift with increasing resolution.
Our theoretical model of the atom-light interaction is in agreement with the observed effect and allows us to quantify its contribution in the uncertainty budget.

\end{abstract}

\submitto{\NJP}
\maketitle

\section{Introduction}

The metastable ${^3\!P_1}$ state in magnesium has a lifetime of 4.4 ms \cite{Hansen08} which is longer than that for heavier alkaline earth elements. This enables us to perform high resolution Ramsey-Bord\'e interferometry \cite{Borde84}, a well established technique in spectroscopy \cite{Ruschewitz98,Udem01,schnatz96,wilpers07}, for interrogation. We apply this method to determine the frequencies of optical transitions from the ground state to the metastable states ${^3\!P_1}$ and ${^3\!P_0}$. The frequency value of the latter transition is derived by combining our measurement for the ${^3\!P_1}$ state with measurements from \cite{Godone93}. This transition is of relevance for the implementation of a lattice clock \cite{Takamoto05}, which is currently underway in our laboratory. Magnesium offers an advantage in that the transition is much less perturbed by blackbody radiation compared to heavier alkaline earth elements currently used for optical clocks \cite{Ludlow08,Rosenband0006}. In strontium, for example, one way to minimize this effect is by performing measurements in a cryogenic environment \cite{Middelmann100}.

The frequency measurement is carried out on an atomic ensemble at a temperature of few milliKelvin and with size comparable to that of the interrogation beams. This makes the measurement very sensitive to phase shifts induced by the wavefront curvature, which are also of concern in atom interferometers \cite{refId0, Kasevich11}. These effects can be minimized by lowering the temperatures, thus yielding more compact clouds, and larger interrogation beam diameters, neither of which were feasible in our case due to high Doppler temperature and laser power constrains. Atom interferometers employ methods such as using interrogation beams in a retro-reflecting configuration \cite{Landragin09}, which cannot be applied to a Ramsey-Bord\'e interferometer used in this work.

The first order Doppler effect caused by the milliKelvin atomic ensemble in combination with the finite wavefront curvature led to a different behavior in our measurement than previously reported by Wilpers, et al. \cite{Wilpers03} in the context of the Calcium atomic clock. 
In our case, the atoms explore a large fraction of the interrogation beam in the transverse direction. This results in atoms on the edge contributing differently, both in terms of phase and excitation probability, to the Ramsey interference pattern than those at the center. We extend the work in \cite{Wilpers03} to develop a theoretical model to account for the effects that occur in our regime. The model is described in section 3.1 in the context of the uncertainty budget of the frequency measurement.

\section{Experimental apparatus}

\subsection{The Ramsey-Bord\'e interferometer}

\begin{figure}
\centering
\includegraphics[width=1\textwidth]{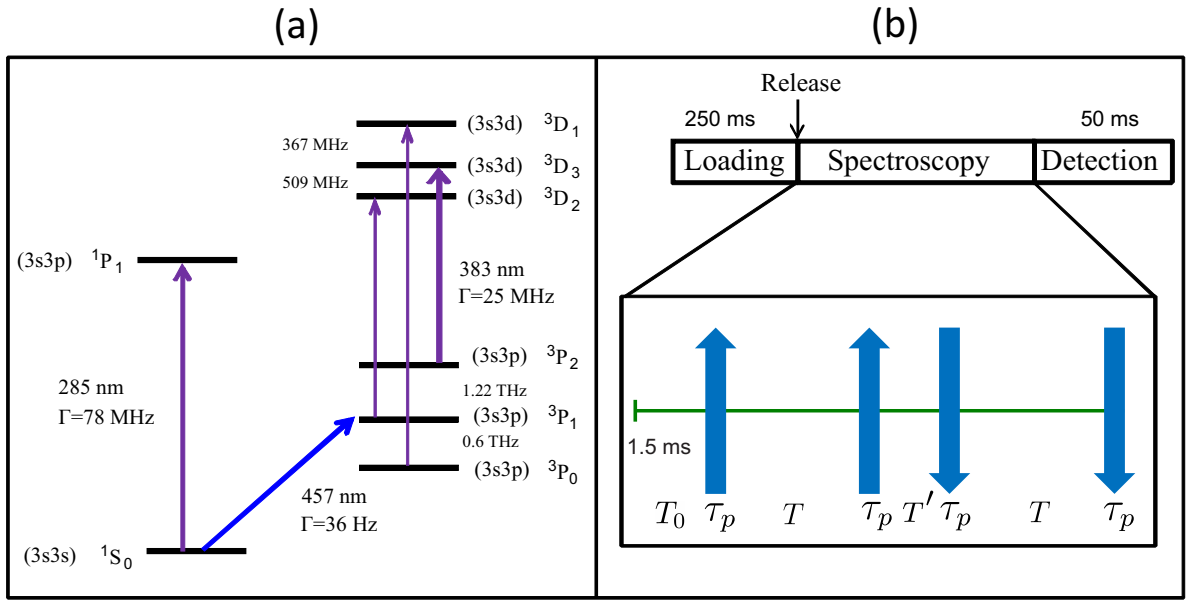}
\caption{{\bf (a)} The electronic states of $^{24}$Mg and the wavelengths of relevant transitions used to manipulate atoms. {\bf (b)} Temporal sequence of the experiment includes preparation, coherent manipulation (spectroscopy) and detection of atoms. The spectroscopy part of the sequence is shown in detail. Arrows indicate the direction of laser beams. $T_{0}$ (=~1.5~ms) is the delay at the start of the interferometer after preparation, $\tau_p$ is the duration of the light pulses and $T^{\prime}$ (=~20~$\mu$s) is the time between the two pairs of light pulses. The coherent evolution time between the first and the second light pulse is $T$. The four pulse Ramsey-Bord\'e interferometer sequence itself lasts for $2T+4\tau_p$+$T^{\prime}\approx$ 0.5~ms. The interferometer is always operated in a symmetric configuration where the delay between the third and fourth light pulse is also $T$. For the frequency measurement, $T$~=~$204~\mu$s and $\tau_p$~=~$3.0~\mu$s.\hfill~}
\label{a:mgtermschema}
\end{figure}

The frequency measurements were performed by locking an ultrastable laser to a time-domain Ramsey-Bord\'e atom interferometer \cite{Borde84,Ruschewitz98}. Figure~\ref{a:mgtermschema}(a) shows the partial $^{24}$Mg level scheme with all relevant transitions. Magnesium is heated in an oven to 420$^\circ$~C to produce an effusive thermal beam. Laser cooling is carried out on the ${^1\!S_0}\rightarrow{^1\!P_1}$ closed transition at 285 nm and the atoms are captured in a magneto optical trap (MOT). The light for driving this transition is produced by a Rhodamine 6G dye laser which outputs 1.2 W of power at 571 nm, followed by cavity enhanced second harmonic generation (SHG) inside a $\beta$-BaB$_2$O$_4$(BBO) crystal which yields 60 mW at 285 nm. Light from the dye laser is transported to the SHG via a polarization maintaining fiber with $\approx$ 50\% efficiency.
The broad linewidth (78~MHz) of the cooling transition implies a relatively short Zeeman slower length of the order of 10 cm. This small distance allows us to use the quadrupole magnetic field of the MOT itself (0.65~T/m along the radial direction) as field for the slowing, avoiding additional coils. The MOT field together with a laser beam carrying 15 mW of power, red detuned by 2.5 full linewidths from resonance and counter propagating to the atomic beam increases the captured atom number by a factor of 25. We load $2.5\times10^7$ atoms in the MOT in 250~ms at a peak density of $4\times 10^{15}$~m$^{-3}$.  
The ensemble has a temperature of 3~mK (which is slightly above the Doppler limit of 1.9~mK) for a laser detuning of one full linewidth below resonance.

After the loading sequence, the cooling light and the MOT magnetic field are turned off within 700~$\mu$s. Atoms are released from the trap at this stage and the cloud expands ballistically in a homogeneous magnetic field of 0.6~mT. This magnetic field is created by large Helmholtz coils and is essential to define the quantization axis allowing the separation of different Zeeman sublevels of the $^3\!P_1$ state. 

\begin{figure}
\centering
\includegraphics[width=1\textwidth]{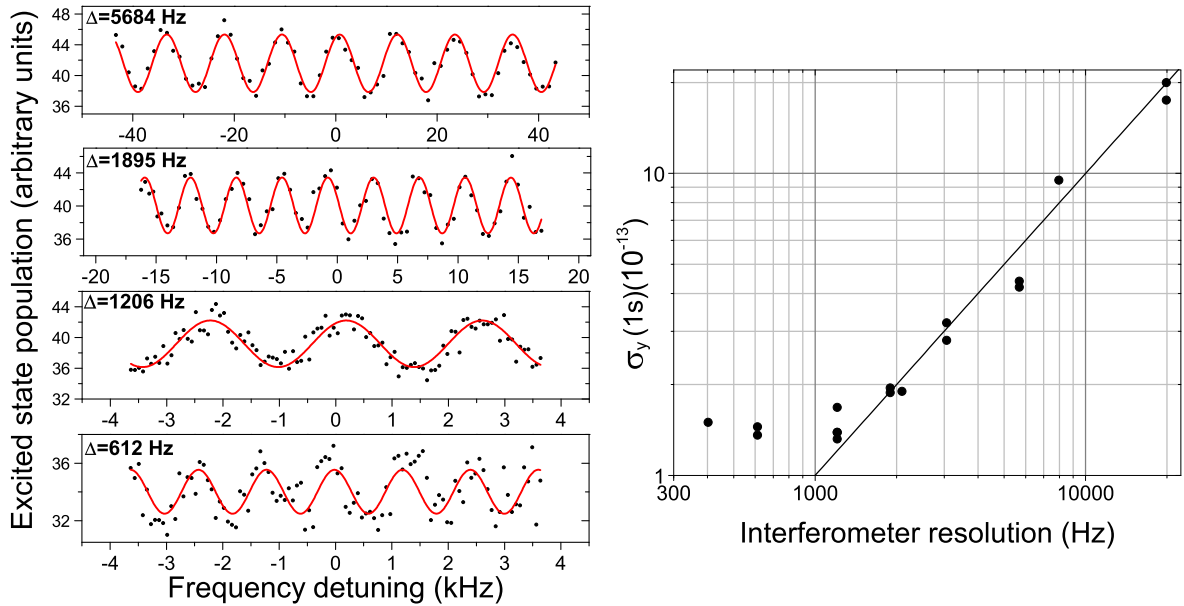}
\caption{{\bf (Left)} Typical interference scans obtained by tuning the laser frequency. The frequency detuning is with respect to the center of the atomic transition. Data points are individual runs without averaging. The fringe period and hence the frequency resolution was adjusted by changing the time $T$. 
$T$ (=~40.6, 128.6, 204, 405 $\mu s$) increases from top to bottom. The pulse duration $\tau_p=3.0~\mu s$ and time between the laser pulse pairs $T^{\prime}=20~\mu s$. {\bf (Right)} Fractional instability of the frequency standard as a function of resolution obtained from the Ramsey pattern according to equation~\ref{StabilityEquation}. The solid line is the theoretical stability for a signal to noise ratio of 3.\hfill~}
\label{a:ramseysundstab}
\end{figure}

The interferometer is operated on the ${^1\!S_0}\rightarrow{^3\!P_1}$ transition and the $m_{J}=0\rightarrow m_{J'}=0$ clock transition is probed. The temporal sequence of the experiment is depicted in figure~\ref{a:mgtermschema}(b).
The interferometer sequence starts 1.5~ms after the release when effects due to the decaying magnetic field are negligible. The peak density drops to $2\times 10^{14}$~m$^{-3}$ due to the expansion. Four $\pi$/2 pulses of duration $\tau_p$ are used to coherently split and recombine the atomic ensemble. The direction of the second pulse pair is reversed with respect to the first one. The time $T'$ between first and second pair of pulses is fixed to 20~$\mu$s. 
The experimental sequence is dominated by the time for preparation and detection of atoms. The evolution of atoms in the Ramsey-Bord\'e interferometer lasts for $2T+4\tau_p$+$T^{\prime}\approx$ 0.5~ms.
The time $T$ between the first and the second pulse and $\tau_p$ both determine the frequency resolution $\Delta$ of the interferometer and is given by $\Delta=1/(4T+4\tau_p)$ \cite{Keupp05}.
The spectroscopy (clock) laser is an amplified diode laser at 914 nm that is stabilized to an ultrastable high finesse resonator \cite{Papepaper}. A fiber frequency comb is used to compare the frequency of the clock laser to a reference frequency source. The stabilized clock laser output is frequency doubled using cavity enhanced SHG in a non-linear KNbO$_3$ crystal and used for interrogation. The interrogation beams are collimated to a 1/e$^2$ radius of 2~mm. 

The large Doppler width of the ensemble (3~MHz), small homogeneous width of the $\pi$/2 pulse (0.3~MHz) and limited power in the interrogation laser (40~mW) limits the fraction of excited atoms to 1\%. In order to detect this small fraction, we measure the excited state population by pumping those atoms to the $^3\!P_2$ state via the $^3\!D_2$ state. The atoms are then captured in a MOT based on the ${^3\!P_2}\rightarrow{^3\!D_3}$ cycling transition in the metastable manifold. The relatively smaller fine structure splitting between the different $^3\!D$ states results in off resonant excitation to $^3\!D_1$ and $^3\!D_2$ states. These states, after spontaneous decay, populate the $^3\!P_0$ and $^3\!P_1$ states. Hence, detection is done in the presence of the repump laser for the $^3\!P_1$ state and an additional repumper for the $^3\!P_0$ state via the $^3\!D_1$ state. The light for these three transitions is produced by amplified diode lasers at 767 nm followed by cavity enhanced SHG in a LiB$_3$O$_5$ (LBO) crystal. The cooling laser is capable of delivering a total of 15 mW at 383 nm to the atoms after losses in the acousto optic modulator (AOM) and fiber coupling. The repump lasers can deliver a total of 2 mW each. With this background free detection scheme, we greatly increase the signal-to-noise ratio in comparison to conventional ground state detection. Flourescence light of the metastable MOT is collected by a lens with a collection efficiency of $10^{-4}$ and imaged on a photomultiplier tube with a quantum efficiency of 25~\%. In order to avoid limitations by photon shot noise, we choose the detection time to be 50~ms giving a typical photon count of 8 per atom. The cycle time of the frequency measurement is 400~ms.

Figure~\ref{a:ramseysundstab} shows Ramsey fringes obtained for typical experimental parameters. All data points were obtained without any averaging. The observed fringes were fitted with sinusoidal functions. The amplitude of the fit gives the signal while the root mean square of residuals are used to determine the noise. The fractional frequency instability at 1~second is estimated using \cite{Riehle},

\begin{equation}
\sigma_y (\tau)=\frac{1}{\pi~Q~\textnormal{S/N}}~\sqrt{T_c/\tau}.
\label{StabilityEquation}
\end{equation}
Here, $\tau$ is the integration time, $Q=\nu_{0}/\Delta$, $\nu_{0}$ is the transition frequency, 2$\Delta$ is the fringe width, $T_c$ is the cycle time and S/N is the signal-to-noise (i.e. fringe amplitude (due to contrast) to rms noise) ratio. Figure~\ref{a:ramseysundstab} also shows the calculated instability at 1 s against the interferometer resolution. The highest stability can be expected for resolutions finer than 1~kHz. For coarser resolutions, the signal-to-noise ratio is constant and the instability increases linearly. In this regime, the stability is limited by fluctuations in the atom number. For finer resolutions below 1~kHz, the fringe contrast drops considerably as is shown in figure~\ref{a:ramseysundstab}. We expect vibrations to be the main cause for this effect as degradation due to wavefront curvature (discussed in the next section) alone is not sufficient to explain this behavior.

The interferometer signal is composed of two Ramsey fringe patterns separated by twice the recoil shift of 39.8~kHz. To obtain the highest contrast, the dark time $T$ is ideally chosen in a way to maximize the overlap of the two Ramsey fringe patterns . This happens at a discrete set of values for $T$. The measurements in figure~\ref{a:ramseysundstab} were  performed only at these specific values. From these measurements we expect the highest stability for the frequency measurement at $T=204.0~\mu$s which corresponding to a resolution of 1.2~kHz. In order to generate an error signal which is insensitive to linear and quadratic background offset of the signal, we used a square wave modulation technique where the excitation probability is probed at four detunings on the steepest slope of the interferogram \cite{Degenhardt05ca}.

\begin{figure}[t]
\centering
\includegraphics[width=0.8\textwidth]{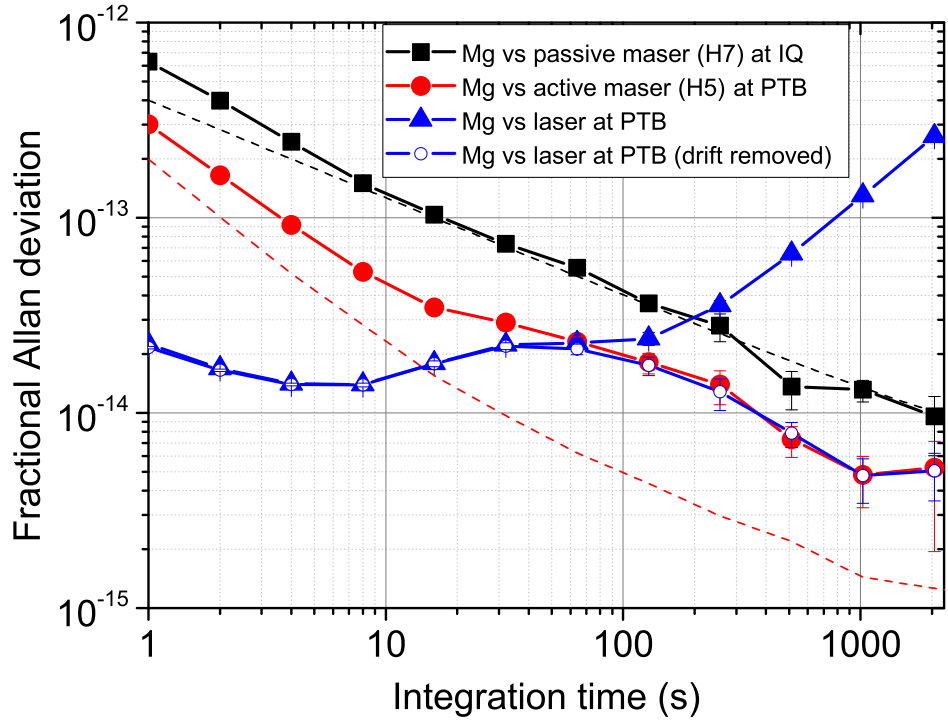}
\caption{Fractional Allan deviation of the frequency of the magnesium standard compared to different references. The black squares show the instability with respect to a local passive hydrogen maser. The red circles (blue triangles) show a remote comparison with respect to an active hydrogen maser (ultrastable laser) at PTB. The dashed black (red) line is the typical stability of the passive (active) maser. 
The best characterization is given by the open blue circles that corrects for the drift in the ultrastable laser for long timescales using the active maser. 
The measurement time covered more than 8000~s allowing us to calculate the Allan deviation for averaging times up to 2000~s. During unlocking events of the dye laser, which in total amounted to less than 250~s, a linear drift correction was applied to the clock laser.\hfill~}
\label{a:StabilityAnalysis}
\end{figure}

\subsection{Remote stability characterization of the frequency standard}

The magnesium frequency standard was characterized with respect to a passive hydrogen maser (H7), stationed locally, and an active hydrogen maser (H5) and the cesium fountain clock CSF1 \cite{Weyers1,Weyers2}, located at PTB, via a phase-stabilized fiber link. Details of the link can be found in \cite{Papepaper}. 
The transfer of stability via a frequency comb from the ultrastable laser to a 1.5~$\mu$m laser can be achieved with an instability of $10^{-15}$ in 1~s and $5\times10^{-18}$ in 10,000~s \cite{Grosche08EPJD}.
The link does not limit the determination of uncertainty of the frequency measurement on any timescale. Even for very long timescales, it does not introduce any uncertainty above the statistical limit \cite{Grosche09OL}. We estimate the fractional uncertainty due to the fiber link to be $<1\times10^{-17}$ for an averaging time of 1000~s.

\begin{figure}[t]
\centering
\includegraphics[width=0.7\textwidth]{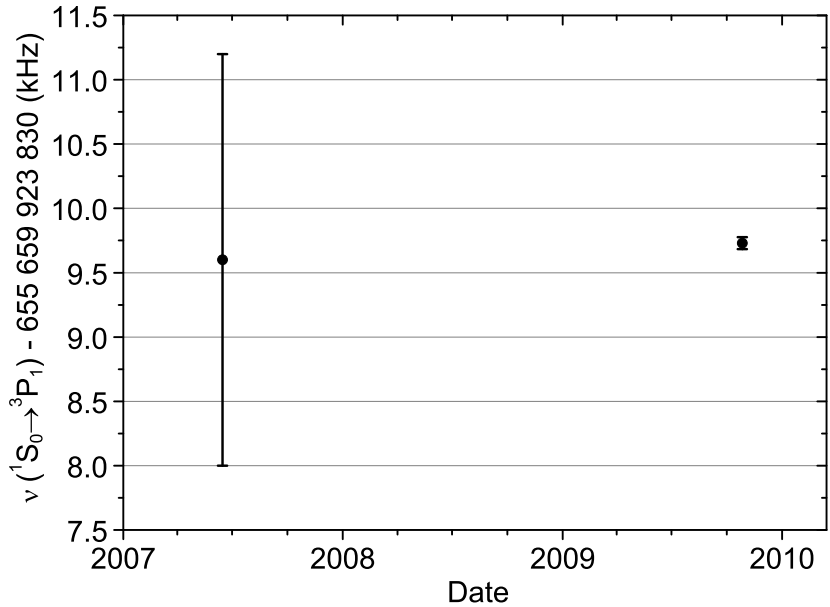}
\caption{The previous \cite{Friebe08} and current frequency measurement of the ${^1\!S_0}\rightarrow{^3\!P_1}$ intercombination transition in magnesium. The 2007 value was obtained with a thermal atomic beam apparatus using a transportable Cs atomic clock of the PTB as reference. 
The current value was obtained by performing spectroscopy on laser cooled atoms with a Cs fountain clock as reference that was stationed remotely at the PTB.}
\label{a:beideFrequenzen}
\end{figure}

The results of the frequency comparisons are shown in figure~\ref{a:StabilityAnalysis}. The local comparison of the clock laser with H7 is dominated by the instability of the microwave system which is $4\times10^{-13}/\sqrt{\tau/\textnormal{s}}$. The comparison with the ultrastable laser at PTB, with a typical stability of few times $10^{-15}$, resolves the short term stability of the magnesium frequency standard. The slight increase in instability above 10~s is due to the tight locking of the clock laser to the atomic resonance. The tight lock allows faster convergence of the laser frequency. Beyond 100~s the instability measurement is dominated by the drifting ultrastable laser at PTB which is not stabilized to an absolute reference. The instability of the magnesium frequency standard for these timescales is resolved with H5.
For averaging times between 100~s and 1000~s, we observe the stability of the magnesium frequency standard to be $2\times10^{-13}/\sqrt{\tau/\textnormal{s}}$. We can study systematic shifts below $10^{-14}$ after an integration time of 500~s .

\section{Frequency measurement}

Several frequency measurements were performed with our frequency standard linked to the fountain CSF1 at PTB. The measurements result in an intercombination transition frequency of  
\begin{equation}
\nu(^{24}\textnormal{Mg:}^1\!S_0 \rightarrow {^3\!P_1})=655~659~923~839~730~(48)~\textnormal{Hz}.
\end{equation}
Figure~\ref{a:beideFrequenzen} shows the new frequency value, which is consistent with our previous measurement \cite{Friebe08}. The applied corrections due to various effects and their uncertainty contributions are summarized in Table~\ref{t:Systematik}. The total uncertainty of 48 Hz amounts to a fractional uncertainty of $7\times10^{-14}$. The individual effects and their uncertainty contributions are discussed in the following sections.

\begin{table}[t]
\centering
\caption{Uncertainty budget for the frequency measurement on laser cooled magnesium atoms.}
\begin{tabular}{lrr}
\hline
\hline
Effect & Frequency correction & Uncertainty \\
\hline
Residual first order Doppler & 0~Hz & 42~Hz \\
AOM phase excursion & 0 Hz & 20~Hz\\
Blackbody radiation & $-0.23$~Hz& $^{+0.04}_{-7.0}$ Hz\\
Quadratic Zeeman &+61 Hz & 7~Hz\\
\hline
Subtotal & $+61$~Hz & 47.6~Hz\\
Statistics of frequency measurement & 0~Hz & 8~Hz\\
Gravitational red shift & $+5$~Hz& 0.4~Hz\\
Fiber link & 0~Hz & 6~mHz\\
Other (from sec 3.2.4) &0 Hz & 2.4~Hz\\
\hline
Total& $+66$~Hz& 48.3~Hz\\
\hline
\hline
\end{tabular}

\label{t:Systematik}
\end{table}

\subsection{Residual Doppler effect}

The relatively high temperature of the freely expanding atomic ensemble results in a significant motion of atoms during the interferometer sequence. This motion causes a residual Doppler shift and manifests itself via two main effects. Firstly, it causes dephasing within the atomic ensemble due to the finite wavefront curvature. Secondly, it modifies the excitation probability of atoms within the ensemble due to the Gaussian intensity profile of the laser beam. Both these effects are exacerbated by the large initial size of the atomic ensemble at the beginning of the interferometry sequence which is of the same order as the interrogation beam diameter.

During each interferometer pulse, an atom samples the local phase and intensity of the light field. Wavefront curvature, beam tilts or an inhomogeneous intensity distribution combined with the movement of the atom between pulses leads to a phase shift in the interferometer. This phase shift enters the interferometric phase via $\Delta\phi=\phi_4-\phi_3+\phi_2-\phi_1$, where, $\phi_i$ denotes the phase of the electromagnetic wave at the position of the atom during the $i^{th}$ interrogation pulse. For phase shifts $\Delta\phi\ll\pi$ this leads to a frequency shift $\Delta\nu=\Delta\phi/(4 \pi (T+\tau_p))$ during the measurement \cite{Wilpers03}. 

We performed a simulation to calculate the frequency shift due to these effects.
The simulation takes into account temperature and size of the atomic cloud, intensity and diameter of the laser beam, wavefront curvature and beam tilts and displacement of the cloud center with respect to the laser beam. The trajectory followed by an individual atom results in a Ramsey fringe with amplitude and phase shift given by the phase and laser intensity at the interrogation pulses, which can be described by a complex amplitude. This complex amplitude when integrated over the entire ensemble results in the overall frequency shift and fringe amplitude. In addition to the analytical treatment of \cite{Wilpers03}, our model considers the different Rabi angles for atoms and is valid for large phase shits. These additional effects are important in our regime and result in a better agreement with the experiment compared to the analytical treatment alone. 
In the following subsections we describe these effects and compare our simulation with experimental data.

\begin{figure}[t]
\centering
\includegraphics[width=0.6\textwidth]{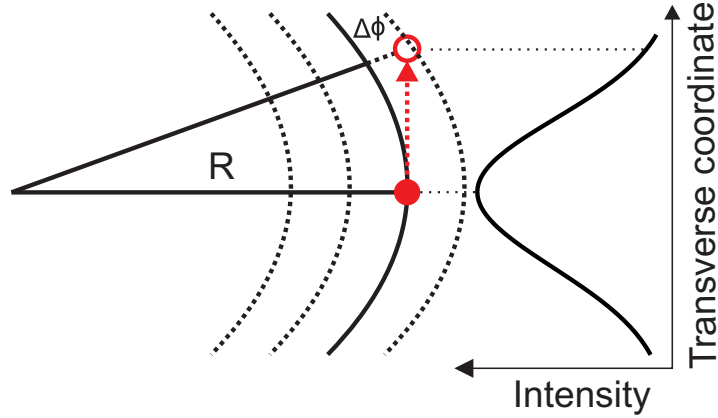}
\caption{Schematic illustrating the position dependent phase and intensity of the light field experienced by an atom due to its transverse motion during the Ramsey-Bord\'e sequence. The interrogation beam wavefront has a radius of curvature R and a Gaussian intensity profile. Filled (open) circle represents the position of the atom during the first (second) pulse. The wavefront curvature causes a phase shift $\Delta\phi$ between the two pulses while the Gaussian intensity profile results in a different Rabi angle at the end of each pulse.\hfill~ } 
\label{a:DephasingCartoon}
\end{figure}

\subsubsection{Dephasing}
 
\begin{figure}
\centering
\includegraphics[width=1\textwidth]{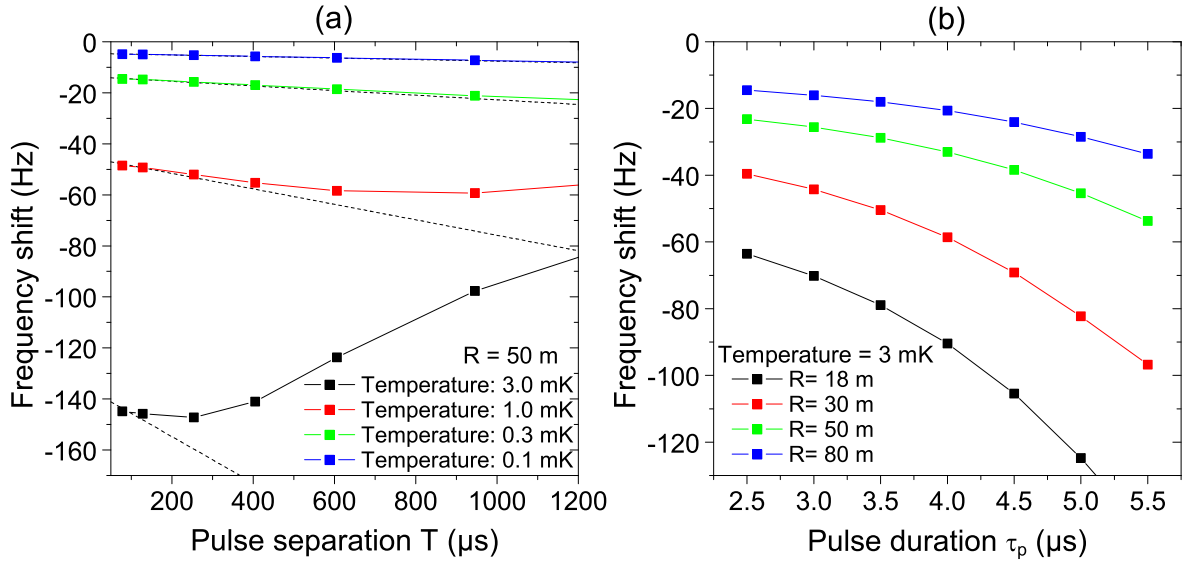}
\caption{{\bf (a)} Calculated frequency shift versus the time delay T in the Ramsey-Bord\'e sequence. The phase accumulated by atoms in the light field depends on their position and velocity. Averaging the phase shifted Ramsey fringes of the individual atomic trajectories leads to a signal with a frequency shift relative to the signal from atoms at rest. In order to illustrate the different effects, we assume a homogeneous intensity profile optimized for a $\pi/2$ pulse but a curved wavefront with a radius of 50~m. Dashed lines are calculations according to the model in \cite{Wilpers03}.
{\bf (b)} Frequency shift vs pulse duration $\tau_p$ for different wavefront curvatures. The temperature of the atoms is 3~mK. Here, we assume an inhomogeneous Gaussian intensity profile of radius w$_{0}$~=~2~mm. 
In both figures, the root-mean-square radius of the ensemble upon release is set to $750~\mu$m and $T_0=1.5$~ms according to experimental values. The calculated dots are connected to serve as guide to the eyes.\hfill~} 
\label{a:Simulations}
\end{figure}

Atoms moving away from the center of a curved wavefront see the phase growing quadratically with distance (figure~\ref{a:DephasingCartoon}). For low temperatures the phase shift remains below $\pi$. Averaged over the entire ensemble this results in a linear increase in frequency shift with respect to T as predicted by the formalism in \cite{Wilpers03} (dashed lines in figure~\ref{a:Simulations}(a)). For higher temperatures, the atoms travel further away from the center and the interferometric phase begins to lie outside the interval [-$\pi$,+$\pi$]. Hence, the Ramsey fringe patterns of atoms with different velocities begin to dephase when the complex fringe amplitudes are averaged over the atomic ensemble.  As fringe patterns from fast moving atoms, that show the largest frequency shift, start to dephase first, they contribute less to the overall interference pattern which is now dominated by the low velocity atoms that remain close to the center of the cloud. Thus the resulting phase shift $\Delta\phi$ is bound and the corresponding frequency shift $\Delta\nu=\Delta\phi/(4 \pi (T+\tau_p))$ is \emph{reduced} with increased T (i.e. for finer resolution). In addition, as T increases the contrast keeps degrading until the interference pattern is completely washed out.

A numerical simulation of the frequency shift due to dephasing with respect to pulse separation time $T$ for laser cooled ensembles of magnesium atoms of different temperature is shown in figure~\ref{a:Simulations}(a). Temperatures of 3 mK, as typically achieved in the experiment, along with moderate curvatures (R~=~50~m) lead to significant frequency shifts of 140~Hz. Only at very short time scales one can treat our scenario with the model in \cite{Wilpers03}. This frequency shift starts to decrease at $400~\mu$s with increasing evolution time T and hence with resolution.
In order to separate the additional effect due to the Gaussian intensity profile of the beams, we assume for this simulation, that the beam diameter is much larger than the atomic ensemble. The radius of curvature of the wavefronts is R~=~50~m and converges along the propagation direction. The two light beams pulsed from opposite directions are assumed to have a similar curvature with the same sign due to our alignment procedure. The effects due to wavefront curvature of the two light fields do not compensate each other in this case. The aforementioned effect will be modified by the Gaussian intensity profile of the laser beam, which will be discussed in the next subsection.

\subsubsection{Effect of the Gaussian intensity profile}

The atoms can spread over a large fraction of the beam diameter, perpendicular to the propagation direction. Atoms on the outer edge of the ensemble are in a low intensity region due to the Gaussian intensity profile of the beam (figure~\ref{a:DephasingCartoon}). This reduces the Rabi angle of these atoms for each pulse and they contribute less to the interference pattern. The power of the beams is unchanged for all pulse times. For longer pulses, atoms in lower intensity regions in the wings have a higher excitation probability and hence contribute more to the ensemble average. The frequency shift thus increases.
Figure~\ref{a:Simulations}(b) shows a simulation of the frequency shift with respect to pulse duration. The power is held constant for different pulse times which results in a different Rabi angle for the different pulse durations. The frequency shift is higher for stronger curvature of the wavefront since atoms experience larger phase changes. 

\subsubsection{Comparison of simulations with experiments}

\begin{figure}
\centering
\includegraphics[width=1\textwidth]{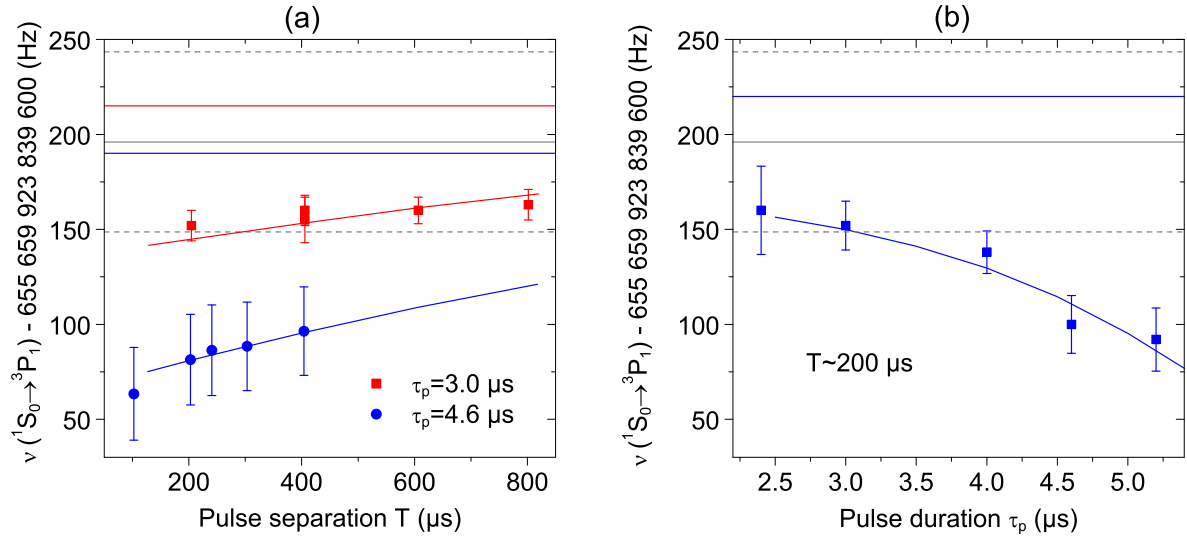}
\caption{
{\bf (a)} Measured frequency versus the pulse separation $T$ between the first and second 
laser pulse of the Ramsey-Bord\'e sequence for two different pulse durations $\tau_p$~=~3.0, 4.6 $\mu$s (red squares and blue circles, respectively). {\bf (b)} Measured frequency versus pulse duration $\tau_p$ for curved wavefronts with $R=18$~m. {\bf Both:} The red and blue curves show the numerically calculated frequency shifts considering the Gaussian shaped laser beam. In order to compare the functional dependence of our model with experimental values, we allowed for a frequency offset to account for systematic shifts such as a mismatch between the atomic cloud center and the optical axis of the laser beams which may cause additional pulse time dependent phase shifts. The red and blue horizontal lines show the unshifted frequencies inferred from our model. The solid gray lines show the mean value of the measurements with optimized wavefronts and the dashed gray lines indicate their uncertainty as derived from the Monte Carlo simulations. The blue data points in (b) have larger error bars because they were measured only with respect to a local passive maser.

\hfill~}
\label{a:experimentdaten}
\end{figure}

\begin{figure}[!t]
\centering
\includegraphics[width=1.0\textwidth]{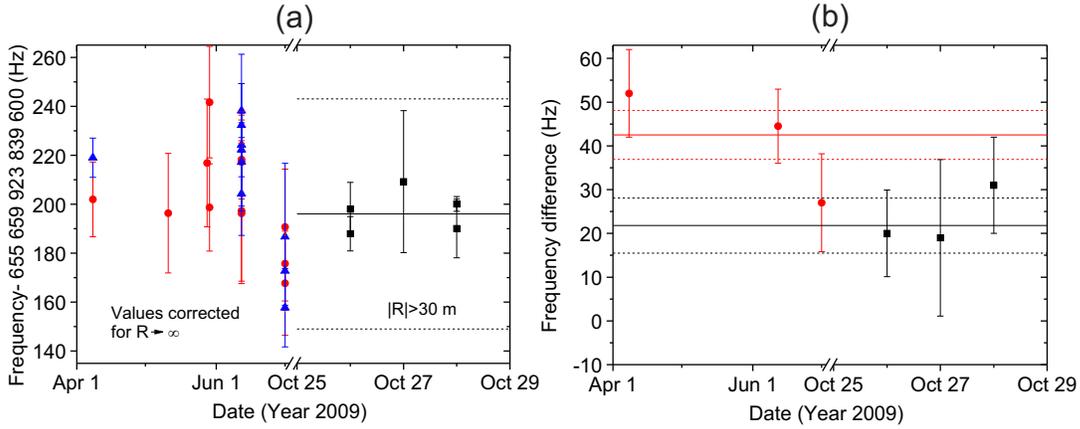}

\caption{{\bf (a)} Measured frequency values of the ${^1\!S_0}\!\rightarrow\!{^3\!P_1}$ transition obtained from two different sets of wavefront curvatures at a resolution of 1.2 kHz. The error bars only account for the statistical errors. All measurements before October 2009 were obtained with curved wavefronts and were corrected according to the theoretical model. Blue and red data points show measurements with a pulse duration of $\tau_p=3.0~\mu$s and $\tau_p=4.6~\mu$s, respectively. Black squares are measurements with optimized wavefronts. For the optimized setting a pulse duration of $3.0~\mu$s was applied corresponding to a Rabi angle of $\pi/2$ for the atoms in the center of the beam. The solid gray line is the weighted average and the dotted line is the statistical uncertainty. 
{\bf (b)} Measured frequency shift for two pulse durations $\tau_p=3.0~\mu$s and $\tau_p=4.6~\mu$s : ($\nu(\tau_p=3.0~\mu s)-\nu(\tau_p=4.6~\mu s)$). Red data points are for curved wavefronts and black data points are for optimized wavefronts. The solid red and gray lines indicate the weighted average and dashed lines the error bars for the respective data sets.
\hfill~}

\label{a:FreqVDate}
\end{figure}

We analyzed the influence of both effects by measuring the phase shift as a function of the resolution and the length of the excitation pulses for a wave front curvature of R = 18~m and compared this with our model. As the two oppositely directed beams of the Ramsey-Bord\'e interferometer were aligned in the same way, they had a similar curvature. This value was determined by measuring the beam radius over a long distance. 

Figure~\ref{a:experimentdaten} shows the results of the first measurement for this relatively strong wavefront curvature. Figure~\ref{a:experimentdaten}(a) depicts the variation of the measured frequency with respect to the pulse separation T (and hence resolution) for two different pulse durations $\tau_p =$~3.0, 4.6 $\mu$s. The measurements confirm that the frequency shift in this regime decreases with increasing resolution. According to our model, the interference pattern's offset phase, which shifts, is very sensitive to the pulse duration. Figure~\ref{a:experimentdaten}(b) presents the frequency values of the intercombination transition obtained for different durations of the excitation pulses.
The different interference scans were performed with a constant intensity. The scans with different pulse duration were performed in random order to minimize other effects such as frequency drifts. 
Pulse duration of $3.0~\mu$s corresponds to a Rabi angle of $\pi/2$ for atoms at the intensity maximum and the Rabi angle scales linearly with pulse duration. 
 
The model correctly predicts the behavior of frequency versus pulse duration for data taken on a single day which gives an increased confidence in our model.
However, the model accepts many parameters that can fluctuate between measurement runs on different days. For example, the frequency values depend critically on the position of the atomic ensemble with respect to the spectroscopy beams for the mentioned wave front curvatures. The frequency can deviate by as much as 5~Hz for a 1~mm misalignment .
Based on our model, we can analyze these fluctuations with a Monte Carlo simulation for different experimental configurations to determine an uncertainty of the absolute frequency shift.
In the Monte Carlo simulation the parameter values such as temperature, position of the ensemble, beam tilts, etc are varied within their estimated uncertainties. This changed the absolute values only by a few percent compared to simulations done with only the expectation values of the parameters. We obtained a correction of +102~Hz and +67~Hz for frequency values obtained with pulse durations of $4.6~\mu$s and $3.0~\mu$s respectively.

Figure~\ref{a:FreqVDate}(a) shows the measured frequency of the intercombination line at a resolution of 1.2~kHz. A correction has been applied to all data points measured with the wavefront of radius R = 18~m. The corrected and uncorrected data points are in good agreement, being a further confirmation of our model. 
The corrected values for curvature of radius R~=~18~m have large error bars and they make only a minor contribution to the final frequency value.
Thus, for the determination of the frequency value of the ${^1\!S_0}\!\rightarrow\!{^3\!P_1}$ transition we only considered the value obtained for R~$>$~30~m. For the optimized wave fronts we still observed a frequency shift for different pulse durations, but it was reduced by a factor of two on average as can be seen in figure~\ref{a:FreqVDate}(b).
We would like to emphasize that our model allowed us to calculate accurately the uncertainty contribution due to the residual Doppler effect. In order to estimate the uncertainty we assume a wavefront curvature of R~=~30~m with an identical sign, which is a conservative lower limit for the magnitude of the curvature. The uncertainty is obtained by performing a Monte Carlo simulation in which we also vary the other experimental parameters within their uncertainty. This result gives a mean frequency shift of 41~Hz with a standard deviation of 7~Hz. Measurement of such a large radius of curvature gave an uncertainty not only in the magnitude but also in the sign of the curvature. This would result in the frequency shift predicted by our model to have the same magnitude but opposite sign. Thus, the correction to the frequency predicted by our model has to be treated not as a frequency shift but as a total uncertainty equal to 42~Hz.

\subsection{Further contributions to the uncertainty budget}
\label{s:furthercontributions}

\subsubsection{AOM phase excursions}

A spurious phase shift due to transient ringing in the excitation pulse results in a frequency shift as demonstrated in \cite{Degenhardt05chirp}. The bandpass of the combined system of switches, RF-amplifiers and the AOM were identified as the reasons for this effect. This shift scales inversely with respect to the pulse duration $\tau_p$ and the pulse separation $T$. From the measurements similar to those shown in figure~\ref{a:experimentdaten}(a), we estimate an uncertainty of 20~Hz in the frequency shift.

\subsubsection{Blackbody radiation}

The blackbody radiation (BBR) due to finite temperature of the surrounding environment shifts the frequency of the intercombination line. A major part of the BBR spectrum does not extend into the optical domain which allows one to treat the field in a quasi-static approximation. The mean square of the electric field due to BBR at temperature $\vartheta$ is $\left\langle E^2\right\rangle=8.55\times10^{-5}~\textnormal{V}^2/(\textnormal{m}^2\textnormal{K}^4)\times\vartheta^4$.
Since the ${^3\!P_1}$ state is anisotropic, the DC polarizability $\alpha$ depends on the direction of the applied electric field. BBR contributes equally from all directions and one has to calculate the weighted average of the differential polarizability which is given by 
$\Delta\alpha=\frac{2}{3}\alpha({^3\!P_{1, m=\pm1}})+\frac{1}{3}\alpha({^3\!P_{1, m=0}})-\alpha({^1\!S_0})=7.3(0.3)~\textnormal{kHz/(kV/cm)}^2$ \cite{Miller19781,Reinsch1976,DissRieger,Bava83,Rieger93}. We estimate the temperature of the science chamber to be 293(10)~K and that of the oven to be 693(10)~K which contributes from a solid angle of only $2\times10^{-3}$. Thus, the necessary correction to the frequency is $-0.23(0.04)$~Hz. However, since we do not know the exact reflectivity of the chamber, we make a worst case approximation and assume that the blackbody radiation from the oven contributes from all directions giving a maximum shift of $-7.2$~Hz. This results in a shift with an asymmetric uncertainty contribution of $-0.23\,(^{+0.04}_{-7.0})$~Hz.
Even for the worst case scenario, the effect is of minor relevance for the uncertainty budget.

\subsubsection{Zeeman effect}
The bias magnetic field applied during spectroscopy along with the earth's magnetic field and the residual quadrupole field from the MOT cause a shift in the clock transition frequency. The largest contribution to this shift comes from the second order Zeeman effect. The coupling coefficient for this interaction is 164~Hz/(mT)$^2$ \cite{Friebe08}.
We determined the value of the magnetic field by measuring the frequency shift of the $m_J=\pm1$ sublevels and obtain values between 11.5~MHz and 14.0~MHz. Using the known splitting of $\pm21$~MHz/mT for the $m_J=\pm1$ states, we calculate the field to be 0.61(12)~mT. This corresponds to a frequency shift of 61(7)~Hz.

\subsubsection{Other contributions}

Further contribution to the uncertainty budget results from a statistical error of the frequency measurement itself and amounts to 8~Hz. Other effects that make minor contributions are stated below. The servo error due to imperfectly compensated resonator drift causes an uncertainty of 2~Hz. 
CSF1 has an instability of $1.4\times10^{-13}$ at 1~s and a systematic uncertainty of $7.6\times10^{-16}$ which corresponds to 0.5~Hz in the optical domain. The statistical uncertainty in comparing the active maser (H5) to CSF1 is 1.2~Hz in the optical domain for a measurement time of 5000~s. The total uncertainty due to the frequency reference is thus 1.3~Hz. 
The AC Stark shift due to the cooling and spectroscopy lasers causes an uncertainty of 0.4~Hz. The gravitational redshift due to difference in height of the magnesium frequency standard and the frequency reference at the PTB is +5(0.4)~Hz. Second order Doppler effect causes a frequency shift of -13(2)~mHz. The fiber link causes an uncertainty of 6~mHz. DC Stark effect, cold collisions and asymmetric background cause insignificant shifts.
The total contribution from these effects keeps the frequency value unchanged while increasing the uncertainty value by 0.2~\%.

\section{Conclusion}

\begin{table}
\caption{Frequency of the strongly forbidden \spzero~transition for different isotopes of magnesium. The values are calculated using results from this work together with data from
 \cite{Godone93, Sterr93, Godone84}.\hfill~}
\vspace{5 pt}
\centering
\begin{tabular}{lrr}
\hline
\hline
Isotope & Frequency (Hz)& Uncertainty (Hz)\\
\hline
$^{24}$Mg & 655~058~646~681~861 & 48\\
$^{25}$Mg & 655~060~050~580~000 & 102~000\\
$^{26}$Mg & 655~061~328~149~000 & 20~000\\
\hline
\hline
\end{tabular}

\label{t:3p0}
\end{table}

We presented the first frequency measurement of the ${^1\!S_0}\!\rightarrow\!{^3\!P_1}$ transition with laser cooled magnesium atoms. The frequency measurement and the characterization was performed by means of PTB's ultrastable and high accuracy frequency references via an optical fiber link. Benefiting from a reduced Doppler shift, we improved the accuracy of the transition frequency by a factor of 30 compared to our previous measurement. As shown in figure~\ref{a:beideFrequenzen}, the new value is consistent with our previous measurement.
Using our accurate spectroscopy measurement and other data in literature, we calculate the frequency of the strongly forbidden $^1\!S_0\!\rightarrow\!{^3\!P_0}$ transition for all magnesium isotopes (Table~\ref{t:3p0}). These transitions are of interest for future optical lattice clocks, one of which is under construction in our group.

\section*{Acknowledgments}
\addcontentsline{toc}{section}{Acknowledgments}
We would like to thank T. Legero for providing the ultrastable reference frequency at 657~nm. We would like to thank B. Lipphardt for help during parts of the experiment. Osama Terra is supported by a scholarship from the Egyptian National Institute of Standards (NIS) and is a member of the Braunschweig International Graduate School of Metrology, IGSM. This joint work was made possible by the framework of the Sonderforschungsbereich 407 and the Centre for Quantum Engineering and Space Time Research (QUEST), with financial support from the Deutsche Forschungsgemeinschaft. T.W. acknowledges the financial support by QUEST.

\section*{References}
\addcontentsline{toc}{section}{References}
\bibliographystyle{unsrt}
\bibliography{onlyone}

\end{document}